\documentclass[12pt]{iopart}
\usepackage[utf8]{inputenc}
\expandafter\let\csname equation*\endcsname\relax
\expandafter\let\csname endequation*\endcsname\relax

\usepackage{amsmath}
\usepackage{amsfonts}
\usepackage{etoolbox}
\usepackage{ulem}
\usepackage{graphicx}
\patchcmd{\subequations}{\alph{equation}}{\textit{\alph{equation}}}{}{}

\usepackage{cleveref}

\crefname{equation}{}{}

\begin{document}

\title{A Solution to the Quantum Measurement Problem}

\author{Z.E. Musielak}

\address{Department of Physics, University of Texas 
at Arlington, Arlington, TX 76019, USA \\}

\begin{abstract}
A novel solution to the quantum measurement 
problem is presented by using a new asymmetric equation 
that is complementary to the Schr\"odinger equation.  
Solved for the hydrogen atom, the new equation describes 
the temporal and spatial evolution of the wavefunction, and
the latter is used to calculate the radial probability density
for different measurements.  The obtained results show 
that Born's position measurement postulates naturally emerge 
from the theory and its first principles.  Experimental verification 
of the theory and its predictions is also proposed.
\end{abstract}


\section{Introduction}

The measurement problem in nonrelativistic quantum mechanics 
(NRQM) has been an active area of research for almost 100 years.  
Numerous attempts to solve it (e.g., [1-29]) have been unsuccessful.
The proposed ideas are aligned to different interpretations of NRQM, 
and they range from the objective wavefunction collapse theories such 
as the Ghirardi-Rimini-Weber (GRW) model [8,9] and its extension 
known as the continuous spontaneous localization (CSL) model 
[10,11] that are based on the Copenhagen interpretation [2-7]], 
to the deterministic DeBroglie-Bohm interpretation with its 
non-local hidden variables [7,15], or to quantum decoherence 
[17-19], and Everett's many-words interpretation [3,15].  Penrose 
[23] proposed a collapse theory that attributes the wavefunction 
collapse to the influence of gravity.  The role of quantum gravity 
in solving the measurement problem was further explored in [24]; 
however, this requires to formulate quantum gravity as a fundamental 
theory of modern physics, which has not yet been accomplished.     

According to the Copenhagen interpretation, quantum measurements 
involve the projection of a previously undetermined quantum state into 
an eigenstate that corresponds to the measurements (e.g., [2,7,15]).  
This cannot be achieved by the Schr\"odinger equation (SE) alone.  
The reason is that the time-evolution of a quantum state $\vert \Psi > 
= \hat  U \vert \Psi_o >$, where $\hat U$ is the time-evolution 
operator for the equation, is continuous and deterministic; however, 
the projection of $\vert \Psi >$ into an eigenstate corresponding to 
the measurements is discontinuous and indeterministic.  Heisenberg 
[25] argued that an energy transfer between a detector and an observed
quantum system disrupts the smooth evolution of the wavefunction 
through quantum jumps (see [26] for a possible resolution]). 

On the other hand, von Neumann [2] tried to solve the measurement 
problem by introducing the reduction postulate, which combines the
SE and the collapse of the wavefunction.  The reduction postulate 
remains controversial [27], and some arguments were made to 
demonstrate that the postulate is incorrect (e.g., [28]) because 
it leads to ambiguity when a measurement must be switched 
on and considered as an interaction [29].  A solution to the 
quantum measurement problem is proposed in this paper.

A quantum system interacting with its surroundings is called 
an open quantum system, and a significant progress in studying 
such systems has been made during the last few decades, with 
the Theory of Open Quantum Systems being established (e.g., 
[30]).  Specifically, theoretical and experimental research on 
open quantum systems by using digital quantum simulators
has been performed and general algorithms for such digital
simulations have been investigated and developed [31-34]. 
However, their practical experimental implementation poses 
some challenges that still remain unsolved [33,34].  The 
approach described in this paper is very different from the 
open system approach, and it has several advantages when 
compared to the latter as the presented theory and its results 
demonstrate.

A solution of the quantum measurement problem proposed
in this paper is based on a new asymmetric equation (NAE) [35], 
whose existence is guaranteed by the irreducible representations 
(irreps) of the Galilean group of the metric [36-38].  The NAE is 
complementary to the SE that also originates from the same irreps 
[39,40].  The main advantage of the NAE, as compared to the SE, 
is that it allows measurements to be directly accounted for.  The 
effects of the measurements on the evolution of the wavefunction 
are investigated by solving the new equation for the hydrogen 
atom.  The obtained solutions represent the temporal and spatial 
evolution of the wavefunction, and they are used to compute the 
radial probability densities of finding the electron after the 
measurement process is completed.

To account for the original electron's probability density 
resulting from the SE, the products of this probablity and 
those resulting from measurements are calculated using 
the NAE.  The obtained products are consistent with 
Born's position measurement postulates (e.g., [41,42]).  
Experimental verification of the calculated probability 
densities is proposed and discussed.  The presented 
results demonstrate that the full description of the 
wavefunction in NRQM may require both the 
Schr\"odinger equation and its complementary 
new aymmetric equation.

The paper is organized as follows: the basic equations resulting 
from the Galilean group are given and discussed in Section 2;
the governing equation is presented in Section 3; the temporal 
and spatial solutions to the governing equation are given in 
Sections 4 and 5, respectively; experimental verification is
discussed in Section 6; and Conclusions are given in Section 7. 

\section{Basic equations from the Galilean group of the metrics}

Galilean space and time are separated and their metrics are given 
by $ds_1^2\ =\ dx^2 + dy^2 + dz^2$ and $ds_2^2\ =\ dt^2$, 
where $x$, $y$ and $z$ are spatial coordinates and $t$ is time.  
The metrics remain invariant with respect to all the transformations 
that form the Galilean group of the metric, ${\mathcal G}$, whose 
structure can be written as ${\mathcal G} \ = S (4) \otimes_s H (6)$, 
where $S(4) = T(1) \otimes R(3)$, with $T(1)$ and $R(3)$ being 
subgroups of translation in time and rotations, respectively.  In addition,
$H(6) = T(3) \otimes B(3)$, where $T(3)$ and $B(3)$ are subgroups 
of translations in 3D space and Galilean boosts, respectively [36-38]. 

The Galilean group of the metrics, ${\mathcal G}$, is a ten-parameter 
Lie group and $H(6)$ is its invariant Abelian subgroup.   The irreducible 
representations (irreps) of $H(6)$ are one-dimensional and they are 
well-knonw [39,40], and they provide labels for all the irreps of the 
entire group ${\mathcal G}$.  Classification of these irreps was done 
by Bargmann [36], who demonstrated that the scalar and spinor irreps 
are physical, but the vector and tensor irreps are not because they do 
not allow for the elementary particle localization.  As a result, in NRQM 
only scalar and spinor wavefunctions are allowed, and their temporal 
and spatial evolution is described by the Schr\"odinger [41,42] and 
L\'evy-Leblond [37,38] equations, respectively.  In this paper, only 
the scalar irreps are considered.  

The transformations for space translations represented by $T(3)$, 
and Galilean boosts represented by $B(3)$, may be considered
separately and written as: $\hat {T}_{\mathbf {a}} \phi(\mathbf {r}, t)\ 
\equiv\ \phi(\mathbf {r} + \mathbf {a}, t)\ = e^{i \mathbf {k} \cdot 
\mathbf {a}} \phi (\mathbf {r}, t)$ and $\hat {B}_{\mathbf {v}} 
\phi(\mathbf {r}, t)\ \equiv\ \phi(\mathbf {r} + \mathbf {v}, t)\ =
\ e^{i \mathbf {k} \cdot \mathbf {v} t} \phi(\mathbf {r}, t)$, 
where $\hat {T}_{\mathbf {a}}$ and $\hat {B}_{\mathbf {v}}$ are 
operators of space translations and boosts, respectively.  Moreover, 
$\phi(\mathbf {r}, t)$ is a scalar wavefunction, $\mathbf {a}$ represents 
a translation in space, $\mathbf {v}$ is the velocity of Galilean boosts, 
and the real vector $\mathbf {k}$ labels the one-dimensional irreps of 
$H(6)$. 

Expanding $\phi(\mathbf {r} + \mathbf {a}, t)$ and $\phi(\mathbf {r} + 
\mathbf {v}t, t)$ in Taylor series and comparing the results to the above
transformations, one obtains the following eigenvalue equation [37,38]
\begin{equation}
- i \nabla \phi(\mathbf {}r, t)\ =\ \mathbf {k} \phi (\mathbf {}r, t)\ ,
\label{eq1a}
\end{equation}

\noindent
which has the same form for the boosts and translations in space, and it 
is the necessary condition that the wavefunction $\phi (\mathbf {}r, t)$ 
transforms as one of the irreps of ${\mathcal G}$.

However, $S(4)$ is not an invariant subgroup of ${\mathcal G}$, which 
means that the above procedure cannot be applied to it; the fact that 
$T(1)$ is an invariant subgroup of $S(4)$ does not help because $T(1)$ 
is not the 'little group' of ${\mathcal G}$ [39].  The proposed solution 
[39] requires that $\phi_{\omega}(\mathbf {r}, t) = \eta(\mathbf {r}, t) 
\phi(\mathbf {r}, t)$, where $\eta(\mathbf {r}, t)$ is a smooth and 
differentiable function to be determined, and $\hat E = i \partial_t = 
i \partial / \partial t$ is the generator of translation; then,  $i \partial_t 
\phi_{\omega}(\mathbf {r}, t)\ =\ \omega \phi_{\omega}(\mathbf {r}, t)$.  
The results presented in [39] demonstrated that $\eta(\mathbf {r}, t) = 
\eta^{\prime}(\mathbf {r}^{\prime}, t^{\prime}) = 1$ in order to be 
in agreement with the Galilean Principle of Relativity.  Thus, the final 
temporal eigenvalue equation [39] can be written as 
\begin{equation}
i \partial_t \phi (\mathbf {r}, t)\ =\ \omega \phi 
(\mathbf {r}, t)\ ,
\label{eq1b}
\end{equation}
and this equation supplements Eq. (\ref{eq1a}).  The eigenvalue equations 
given by Eqs (\ref{eq1a}) and (\ref{eq1b}) guarantee that the wavefunction 
$\phi (\mathbf {r}, t)$ transforms as one of the irreps of ${\mathcal G}$, 
and that these equations can be used to derive dynamical equations that 
are consistent with the Galilean group of the metric ${\mathcal G}$.  

There is another approach to find dynamical equations by using the 
Casimir operator of ${\mathcal G}$, which is well-known (e.g., [39]).
However, this operator does not connect Galilean space and time,
and therefore, it cannot be used to obtain any dynamical equation.
The problem is resolved by extending ${\mathcal G}$, so that the 
connection is allowed.  This results in the so-called extended Galilean 
group, whose structure is $\mathcal{G}_e\ = [R(1) \otimes B(3)] 
\otimes_s [T(3+1) \otimes U(1)]$, where $T(3+1)$ is an 
invariant subgroup of $\mathcal{G}_e$ of translations in space 
and time, and $U(1)$ is a one-parameter unitary group (e.g., 
[37,38]); note that the structure of this group is similar to the 
Poincar\'e group (e.g., [43-46]), and that $\mathcal{G}_e$ has 
three Casimir operators.  If a scalar wavefunction is considered, 
then, one of these Casimir operators gives directly the Schr\"odinger 
equation [39], but this requires a prior knowledge of $\mathcal{G}_e$,
which was introduced only after the Schr\"odinger equation was 
already established and used in NRQM.  Moreover, it must be also 
noted that $\mathcal{G}_e$ is not the group of the Galilean metrics. 
Therefore, the derivation of the Schr\"odinger equation from this Casimir 
operator may not be consistent with the Principle of Galilean Relativity. 

As a result, the eigenvalue equations given by Eqs (\ref{eq1a}) 
and (\ref{eq1b}) are used here to obtain two second-order partial 
differential equations that are asymmetric in space and time 
derivatives.  The coefficients in these equations are expressed 
as ratios of the eigenvalues, which label the irreps of $\mathcal{G}_e$.
If these labels are identified as wave frequencies and wavenumbers, 
the resulting asymmetric equations describe propagation of classical 
waves [47].  However, the coefficients can be determined by using 
the de Broglie relationship (e.g., [41,42]), then, one of the resulting 
asymmetric equations is the Schr\"odinger equation  
\begin{equation}
\left [ i {{\partial} \over {\partial t}} +  \frac{\hbar}{2 m}
\nabla^{2} \right ] \phi_S (t, \mathbf {x}) = 0\ ,
\label{eq1}
\end{equation}  
and the other can be written as 
\begin{equation}
\left [ \frac{i}{\omega} {{\partial^2} \over {\partial t^2}} 
+ \frac{\hbar}{2 m}{\mathbf k} \cdot \nabla \right ] \phi_A 
(t, \mathbf {x}) = 0\ ,
\label{eq2}
\end{equation}  
which is a new asymmetric equation [35], with $\phi_S 
(t, \mathbf {x})$ and $\phi_A (t, \mathbf {x})$ being the 
wavefunctions satisfying these equations.  The equations 
can also be derived from the relativistic energy-momentum 
relation by taking the nonrelativistic limit, which shows 
tha the expression for the nonrelativistic kinetic energy 
$E_k = p^2 / 2m$ underlies both equations.  Based
on the mathematical structure and physical meaning of 
the SE and NAE, the equations are complementary to 
each other, and they describe different aspects of the 
evolution of the wavefunction in NRQM.

There are several important differences between the 
SE and NAE.  The most prominent one is the difference 
in time derivatives.  By being first-order in time, the time 
solutions to the SE are also the solutions to the continuity 
equation for probability, which is also first-order in time.  
This means that the SE is invariant with respect to unitary 
transformations, and that NRQM based on the SE is a unitary 
theory.  On the other hand, the NAE being second-order in 
time may have solutions that do not satisfy the continuity 
equation and, as a result, the NAE may be used to describe 
some non-unitary processes in NRQM.  Thus, the SE (unitary) 
$\rightarrow$ the NAE (non-unitary), with the latter describing 
processes such as the quantum measurement problem 
investigated in this paper, or quantum jumps that represent 
absorption and emission of electromagnetic (EM) radiation 
by atoms, and were recently studied in [47] by using the NAE.

Another difference between the SE and NAE is the presence 
of the eigenvalues $\omega$ and $\mathbf {k}$ in the latter. 
To describe any physical event or process by the NAE requires
specifying both eigenvalues.  By identifying $\omega$ and 
$\mathbf {k}$ as the characteristic frequency and wavevector 
of dark matter, some attempts were made to formulate a 
quantum theory of cold dark matter [48] and galactic cold 
dark matter halos [49]; however, the formulated theories 
require observational verification of their predictions.  In 
this paper, the eigenvalues are identified with measurements 
performed on a quantum system, which is a hydrogen atom 
in its basic 1s state.  The changes in the temporal behavior 
of the wavefunction, and in the radial probability densities 
of the electron resulting from the performed measurements 
are presented and discussed.  

\section{Governing equation and measurements}

The parameters $\omega$ and $\mathbf {k}$ that appear 
explicitly in the NAE (see Eq. \ref{eq2}) are the eigenvalues of 
the Hermitian operators in the eigenvalue equations given by 
Eqs (\ref{eq1a}) and (\ref{eq1b}), which means that they are 
real and label the irreps of the group.  Thus, they can also be 
associated with a measuring apparatus, specifically with the EM 
radiation that this apparatus uses to probe a quantum system.  
For the theory developed in this paper, it is sufficient to prescribe
the frequency of EM radiation $\omega = \omega _{ap}$, which 
gives $E_{ap} = \hbar \omega_{ap}$, and calculate the resulting 
wavevector ${\mathbf {k}} = {\mathbf {k _{ap}}}$, with 
${\mathbf {k _{ap}}} = {\hat k ( 1 / \lambda _{ap}})$, where 
$\lambda_{ap}$ is the wavelength of EM used by the apparatus.
In general, $\omega$ and $\mathbf {k}$ may also be determined 
by temporal and spatial scales associated with the measuring process.

Then, the eigenvalue equations for the energy, $\hat E = i \hbar 
\partial_t$, and momentum, $\mathbf {\hat P} = - i \hbar \nabla$, 
operators acting on the wavefunction $\phi_A$ are given by $\hat E 
\phi_A = \sqrt{E_{ap} E}\ \phi_A$ and $(\hbar \mathbf {k_{ap}} 
\cdot \mathbf {\hat P}) \phi_A = (\mathbf {p_{ap}} \cdot 
\mathbf {\hat P}) \phi_A = p_{ap} p\ \phi_A$, with $E = \hbar 
\omega$, $p = \hbar k$, and $p_{ap} = \hbar k_{ap}$ [47-49].

The new asymmetric equation in the spherical variables, 
with the $r-$dependence only, the potential $V(r)$, and 
the effects of measurements accounted for by the presence 
of $\omega _{ap}$ and ${\mathbf {k _{ap}}}$, becomes 
\begin{equation}
\left [ \frac{i \hbar}{\omega _{ap}} \left ( {{\partial^2} \over 
{\partial t^2}} \right ) + \frac{\hbar^2}{2 m} ( {\mathbf k _{ap}} 
\cdot \nabla ) + V(r) \right ] \phi_A (t, \mathbf {r}) = 0\ ,
\label{eq3}
\end{equation}  
which is complementary to the SE with the potential to be 
specified [40,41]. 

The Schr\"odinger equation with the Coulomb potential,
$V(r) = - e^2 / (4 \pi \epsilon_o r)$, describes a hydrogen 
atom by giving the solutions for the wavefunction in terms 
of Laguerre's polynomials (the radial-dependence) and 
spherical harmonics (the angular-dependence), and predicts
correctly its quantum energy levels given by $E_n = - (m / 2 
n^2 \hbar^2) ( e^2 / 4 \pi \epsilon_o )^2$.  The results 
given by the SE descibe the electron's unitary evolution, 
and they are valid prior to making any measurement. 

As pointed out in Introduction, all previous attempts to 
account for measurements by using the SE have failed,
including von Neumann's approach with the reduction 
postulate because of its ambiguity.  To remove this 
ambiguity, it is suggested herein that at the moment 
when a measuring apparatus is applied to a quantum 
system, the system is described by the NAE.  To verify 
this suggestion, Eq. (\ref{eq3}) is now solved for a 
hydrogen atom in its basic 1s state, and the temporal 
and spatial behavior of the wavefunction resulting from 
the measurement process is calculated.  Then, the spatial 
solutions are used to compute changes in the radial 
probabilty density caused by measurements.  

However, the probability densities resulting from the 
measurements, and calculated with the NAE, do not 
account for the previous electron's unitary dynamics.  
To take this into consideration, the probabilities resulting 
from the measurements are multipled by the original 
probablility density obtained for the electron from the 
SE.  The resulting probabliltes demonstrate that 
the original state of the electron is projected into the 
observable eigenstate, as suggested by Born [50,51]; 
this is now known as Born's position measurement 
postulates or Born's rules. 

Born's rules suggest that when a quantum particle 
interacts with a measuring apparatus, then it gets confined
into a measuring eigenstate that corresponds to a well-defined
position (e.g., [41,42]).  In other words, the rules connect 
NRQM and its theoretical framework to experiments because 
they allow calculating probablilities from quantum mechanical 
amplitudes.  Thus, they become a crucial link between abstract 
NRQM and the real world of experience.  However, no first 
principle derivations of the rules exist so far, and they are 
considered to be Born's intelligent guess (e.g., 52,53]).  In 
this paper, Born's rules are described by the NAE, which 
makes them to originate from the same basic principles 
as the NAE does, namely, from the irreps of the Galilean 
group of the metric and the Principles of Galilean Relativity.

\section{Time-dependent solutions and their physical meaning} 

To solve Eq. (\ref{eq3}), the spherical variables are separated 
into the temporal and spatial (radial only) components, $\phi_A 
(t, \mathbf {r}) = \chi (t)\ \eta (\mathbf {r})$, and $- \mu^2$ 
is chosen as the separation constant.  To relate this approach
directly to the hydrogen atom problem and its description 
by the SE, it is assumed that $V(r)$ is the Coulomb 
potential and that $- \mu^2 = E_n$, which gives     
\begin{equation}
{{d^2 \chi} \over {d t^2}} + i \left  ( \frac{E _{ap}}{2 m} \right )
\left ( \frac{1}{n a_o} \right )^2 \chi = 0\ ,
\label{eq4}
\end{equation}  
and
\begin{equation}
{{d \eta} \over {d r}} + \frac{1}{(\mathbf {\hat {k} _{ap}} 
\cdot \mathbf {\hat {r}}) k_{ap}\ a_o} \left  ( \frac{1}{n^2 
a_o} - \frac{2}{r} \right ) \eta = 0\ ,
\label{eq5}
\end{equation}  
where $E_{ap} = \hbar \omega _{ap}$, $\mathbf {k_{ap}} = 
k_{ap} \mathbf {\hat k}_{ap}$, $\mathbf {r} = r \mathbf 
{\hat r}$, and the Bohr radius $a_o = (4 \pi \epsilon_o 
\hbar^2) / m e^2 = \hbar / (m c \alpha)$, with $\alpha$ 
being the fine structure constant.  It must be noted that both 
$\omega _{ap}$ and $k _{ap}$ are specified by interactions 
between a measuring apparatus and a quantum system.

Using $i = (1/ \sqrt{2} + i / \sqrt{2})^2$, the solutions to Eq. 
(\ref{eq4}) are 
\begin{equation}
\chi (t) = C_{\pm} \exp{ \left [ \pm\ i \left ( \frac{1}{\sqrt{2}} 
+ \frac{i}{\sqrt{2}} \right ) \sqrt{\frac{E _{ap}} {2m}} \frac{t}
{n a_o} \right ]}\ ,
\label{eq6}
\end{equation}  
where $C_{\pm}$ are the integration constants corresponding 
to the $\pm$ solutions.  The solutions require that a measuring 
apparatus interacts with a quantum system, that is $\omega_{ap} 
\neq 0$ as otherwise $\chi = C_{\pm}$, and the temporal 
evolution of the wavefunction must be described by the SE.  
Both solutions with $C_{+}$ and $C_{-}$ are physical and 
they correspond to $t \rightarrow + \infty$ and $t \rightarrow 
- \infty$, respectively, which shows that the solutions are 
symmetric with respect to $t = 0$.  In the following, only 
the solution with $C_{+}$ is considered because $t = 0$ 
is the intital time at which measurement begins, and $t > 0$ 
when it is performed.

The real part of the solution is 
\begin{equation}
{\cal {R} \it e} [\chi (t)] = C_{+} \cos{ \left ( \sqrt{\frac
{E _{ap}}{m}} \frac{t}{2 n a_o} \right )}  \exp{ \left ( - 
\sqrt{\frac{E _{ap}}{m}} \frac{t} {2 n a_o} \right )}\ .
\label{eq7}
\end{equation}  
The arguments of the cosine and exponential functions
can be estimated when $\omega_{ap}$ is given by 
specifying a measuring apparatus.  Consider a quantum 
microscope designed to measure the orbital structure 
of the hydrogen atom (e.g., [54]).  Then, $\omega_{ap}$ 
is the frequency of electromagnetic (EM) waves used to 
perform the measurement.  Let this frequency be either 
the optical, ultraviolet, or X-ray part of the EM spectrum, 
then the factor $(\sqrt{E _{ap} / m} / a_o)$ in the 
arguments of the cosine and exponential functions 
becomes either $\sim 10^{15}$ $s^{-1}$, $\sim 
10^{16}$ $s^{-1}$, or $\sim 10^{17}$ $s^{-1}$, 
respectively.  As a result, the arguments are very large 
and the exponential function decays very rapidly for 
any $t > 0$, while the cosine function remains periodic 
despite its large argument.  The time of this decay of 
the wavefunction decreases with increasing frequency 
of the used EM waves.

Since the argument of the exponential function is very large,
the temporal part of the wavefunction reaches zero on a very 
short time scale that ranges from $\sim 10^{-17}$ $s$ to 
$\sim 10^{-15}$ $s$.  The rapid decay of the wavefunction 
in time predicted by the NAE and its temporal solutions may 
be identified as the duration of the measurement after which 
the electron resumes its original orbital.  If the decay is considered
'instantaneous', then ${\cal {R} \it e} [\chi (t)]$ may represent 
the wavefunction that impinges on a screen or photographic 
film in experiments with double-slits, or diffraction of electrons 
through a narrow aperture, and be identified with the 
'instantaneous collapse' of the wavefunction (e.g., [55]).  

In general, it is expected that the time required for the 
wavefunction collapse is shorter when $m$ increases (e.g., 
[7,15,41,42]).  The arguments of the periodic and exponential 
parts of the solution ${\cal {R} \it e} [\chi (t)]$ are the same, 
and they depend explictly on the mass of electron $m$.  Thus, 
with $a_o$ being a function of $m$, the arguments depend 
on $\sqrt{m}$, whose larger value causes the exponential 
function to decrease faster, and shortens the time of the 
wavefunction collapse.

\section{Time-indepedent solutions and probability densities}

Having obtained the solutions for the temporal evolution
of the wavefunction, the solutions for its spatial behavior 
are now derived, and the resulting probabilty densities 
are calculated.  Defining $\beta _{ap} = [(\mathbf {\hat 
{k} _{ap}} \cdot \mathbf {\hat {r}}) k _{ap}\ a_o]^{-1}$, 
the solution to Eq. (\ref{eq5}) can be written as 
\begin{equation}
\eta (r) = \eta_{0}\ r^{2 \beta _{ap}}\ \exp{ \left [ - \left ( 
\frac{\beta _{ap}}{n^2} \right ) \frac{r}{a_o} \right ]}\ ,
\label{eq8}
\end{equation}  
where the integration constant $\eta_0$ normalizes the 
wavefunction and is evaluated from  
\begin{equation}
4 \pi \int_0^{\infty} r^2 \vert \eta (r) \vert^2 dr = 1\ ,
\label{eq9}
\end{equation}  
showing that $\eta_0$ depends on $\beta_{ap}$ and 
on the principal quantum numbers $n$; thus, $\eta_o 
= \eta_{o,n} (\beta_{ap})$.  Moreover, in spherical 
symmetry $\mathbf {\hat r}$ can always be aligned 
with $\mathbf {\hat {k} _{ap}}$, which means that 
$\mathbf {\hat {k} _{ap}} \cdot \mathbf {\hat {r}} 
= 1$ and $\beta _{ap} = (k_{ap}\ a_o)^{-1} = 
\lambda_{ap} / a_o$, where $\lambda_{ap}$ is the 
wavelength of EM waves used in the measuring process.  

The radial probability density in a spherical shell volume 
element is given by   
\begin{equation}
{\cal{P}}_{n} (r) = r_{max} \frac{dP (r)}{d r}  = 
\eta^2_{o} (\beta_{ap}) \left ( \frac{r}{r_{max}} 
e^{ - r / r_{max}} \right )^{4 \beta_{ap} +2}
\label{eq10}
\end{equation}  
where $\eta^2_{o} (\beta_{ap}) = 1 / I (\beta_{ap})$.
Defining $x =  r / r_{max}$, the integral $ I (\beta_{ap})$
becomes 
\begin{equation}
 I (\beta_{ap}) = \int_{0}^{\infty} ( x\ e^{-x})^{4 
\beta_{ap} +2}\ dx\ ,
\label{eq11}
\end{equation}  
and it can be evaluated after $\beta_{ap}$ is specified. Moreover, 
$r_{max}$ is the most probable radius defined as 
\begin{equation}
r_{max} = n^2 a_o \left ( 2 + \frac{1}{\beta_{ap}} \right ) =
n^2 a_o \left ( 2 + \frac{a_{o}}{\lambda_{ap}} \right )\ .
\label{eq12}
\end{equation}  
The effects of measurements are given by $\beta_{ap} \neq 0$ 
(or $\lambda_{a} \neq 0$), which is required for the NAE to 
represent the radial wavefunction and its evolution.  In the 
absence of measurements, the wavefunction's evolution 
is described by the SE.   

To compare the obtained results directly to the radial probability 
densities of a hydrogen atom given by the time-independent 
solutions to the SE, it is convenient to introduce $r_a = r / a_o$.
Then, the radial probability density ${\cal{P}}_{n} (r_a)$ given 
by Eq. (\ref{eq10}) can be written as
\begin{equation}
{\cal{P}}_{n} (r_a) = \frac{dP (r_a)}{d r_a}  = \eta^2_{o,n} 
(\beta_{ap})\ r_a^{4 \beta_{ap} +2}\ e^{-2 (\beta_{ap} / n^2) 
r_a}\ ,
\label{eq13}
\end{equation}  
where $\eta^2_{o,n} (\beta_{ap}) = 1 / I_n (\beta_{ap})$, and 
the integral $I_n (\beta_{ap})$ can be evaluated by using 
\begin{equation}
I_n (\beta_{ap}) = \int_{0}^{\infty} r_a^{4 \beta_{ap} +2}\ 
e^{-2 (\beta_{ap} / n^2) r_a} = \frac{1}{(2 
\beta_{ap}/n^2)^{4 \beta_{ap} +3}}\ \Gamma (
{4 \beta_{ap} +3})\ ,
\label{eq14}
\end{equation}  
which is valid if ${\cal {R} \it e} (2 \beta_{ap}) > 0$ and 
${\cal {R} \it e} (4 \beta_{ap} + 2) > -1$; note that both 
conditions are obeyed in the theory presented in this paper.\\
%

%
\begin{figure}
\begin{center}
\includegraphics[width=120mm]{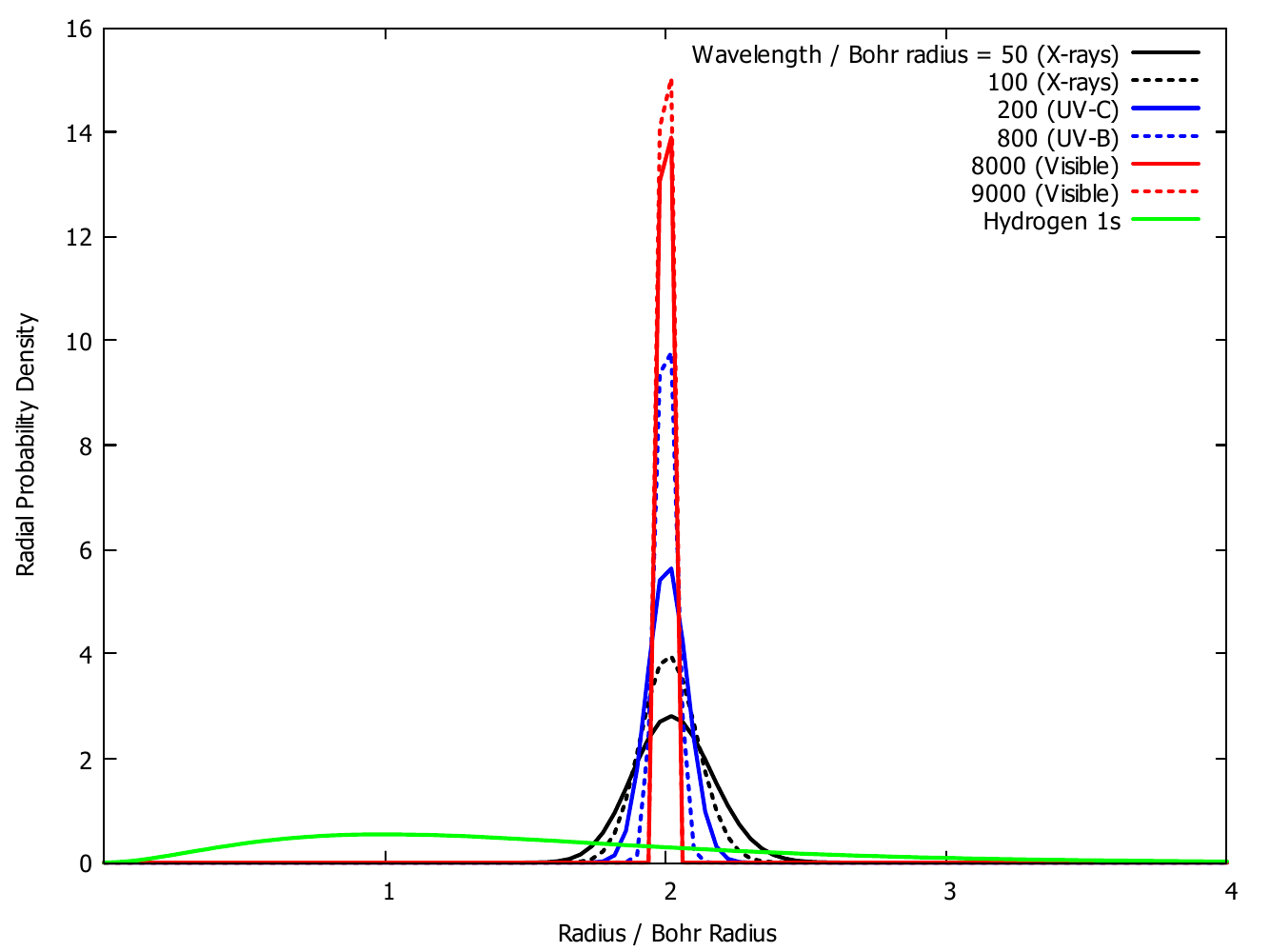}
\caption{The radial probability density, ${\cal{P}}_{1} (r_a)$, 
is plotted versus the ratio of radius to the Bohr radius, $r / a_o$,
for $n = 1$ and $\beta_{ap} = \lambda_{ap} / a_o$ that 
ranges from X-rays to the ultraviolet and visible parts of the 
EM spectrum.  The theoretically predicted ${\cal{P}}_{1} (r_a)$ 
is compared to the radial probability density, ${\cal{P}}_{1s} 
(r_a)$, for the hydrogen $1s$ orbital.}
\label{fig.1}
\end{center}
\end{figure}
%

Using Eqs (\ref{eq13}) and (\ref{eq14}), the radial probability 
density ${\cal{P}}_{1} (r_a)$ is calculated for $n = 1$ and 
for different values of $\beta_{ap}$, which varies from $50$ 
to $9000$.  This range of $\beta_{ap}$ corresponds to 
$\lambda_{ap} = 50\ a_o$ and $100\ a_o$ (X-rays), 
$\lambda_{ap} = 200\ a_o$ and $800\ a_o$ (ultraviolet), 
and $\lambda_{ap} = 8000\  a_o$ and $9000\ a_o$ (visible); 
this range can be extended to the other parts of the EM spectrum.  
Since the selected values of $\beta_{ap}$ are positive integers, 
the function $\Gamma ({4 \beta_{ap} +3}) = ({4 \beta_{ap} 
+ 2})!$ in Eq. (\ref{eq14}).  The obtained results are presented 
in Fig. 1 showing also the radial probability density, ${\cal{P}}_{1s} 
(r_a) = 4 r_a^2\ e^{-2 r_a}$, for the hydrogen $1s$ orbital 
plotted for comparison.  

The maxima of ${\cal{P}}_{1} (r_a)$ plotted in Fig. 1 are 
at $r = r_{max}$, where the latter depends on $\beta_{ap}$ 
(see Eq. \ref{eq12}).  This dependence on $\beta_{ap}$ or
$\lambda_{ap}$ is reflected in Fig. 1 as small shifts towards 
the larger values of $r_a$ for shorter wavelengths, such as 
X-rays and UV.  However, in the limit $\lambda_{ap} 
\rightarrow \infty$, one finds $r = r_{max} = 2 a_o$, and
\begin{equation}
{\cal{P}}_{1} (r_a) = \lim_{\lambda_{ap} \rightarrow \infty} 
\eta^2_{o,1} (\lambda_{ap}) r_a^{4 \lambda_{ap}/a_o +2}\ 
e^{-2 (\lambda_{ap}/a_o) r_a} = \delta (r - 2 a_o)\ ,
\label{eq15}
\end{equation}
where $\delta (r - 2a_o)$ is the Dirac delta function.  This confirms
that for very long wavelengths, ${\cal{P}}_{1} ( r = 2 r_a) = 1$, 
which is the classical limit of the measuring process.  The main 
result is that a quantum particle interacting with a measuring 
apparatus gets confined into a measurement eigenstate, or in 
a well-defined position at $r = 2a_o$ as originally suggested 
by Born's position measurement principles (e.g., [41,42]). 

The theory and results presented in this paper descibe the 
process known as quantum decoherence, which occurs when 
a quantum  system interacts with its environment.  As a result 
of this interaction, the information that the system contains is 
admixed up with its environment.  More specifically, the system 
loses its ability to show superposition and interference effects, 
and its physical behavior resembles a classical system.  This can 
be seen in Fig. 2, which shows that the particle is confined to a 
very narrow eigenstate after measurements are done.  Moreover, 
according to Eq. (\ref{eq15}), the longer the wavelength of EM 
radiation absorbed by the system, the more prominent its 
classical-like behavior is observed.
%

%
\begin{figure}
\begin{center}
\includegraphics[width=120mm]{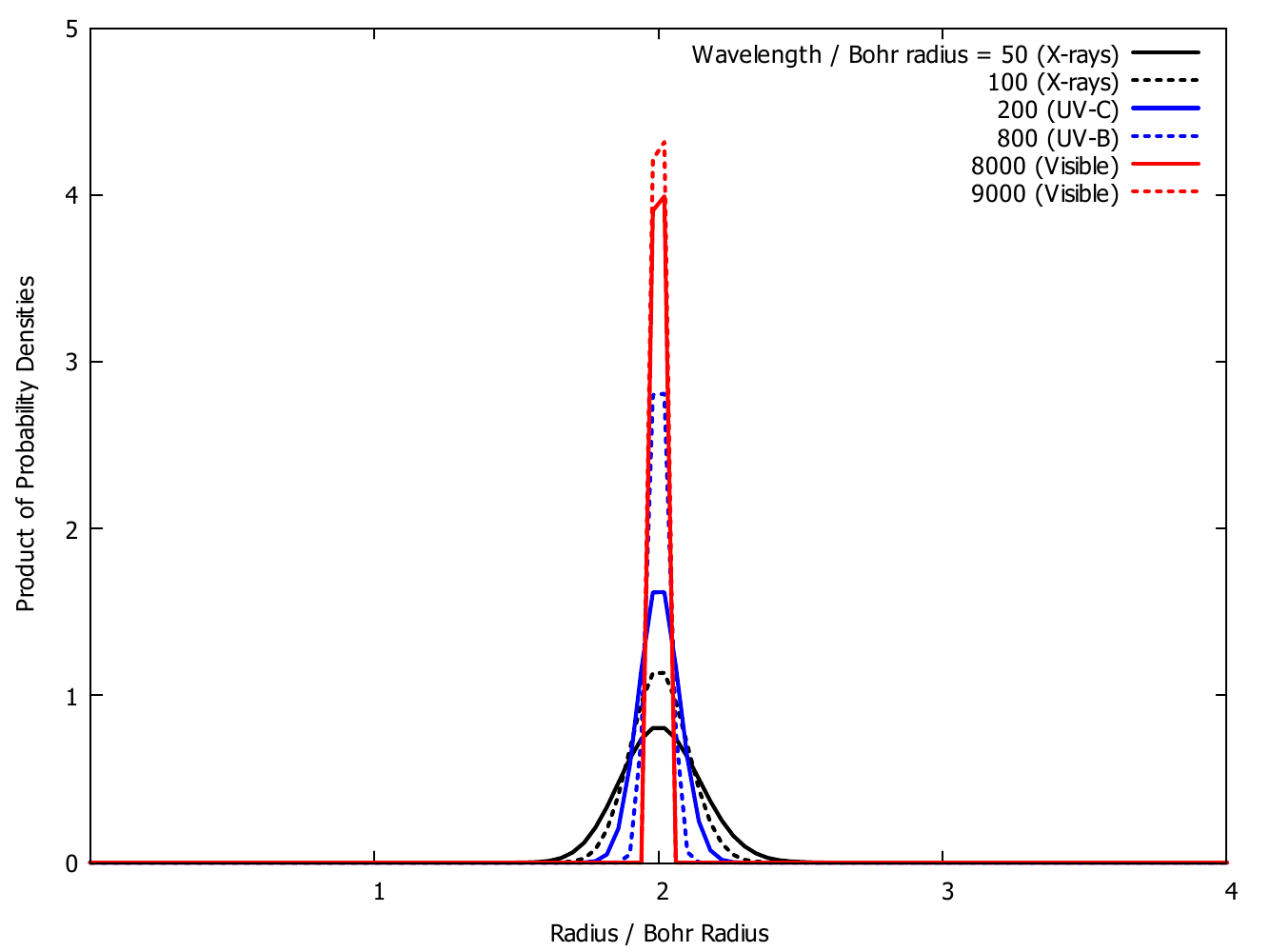}
\caption{The product of the radial probability densities, 
${\mathfrak{P}} (r_a) = {\cal{P}}_{1} (r_a) \times 
{\cal{P}}_{1s} (r_a)$, plotted versus the ratio of 
radius to the Bohr radius, $r / a_o$, for  $n = 1$, 
and $\beta_{ap} = \lambda_{ap} / a_o$that ranges 
from X-rays to the visible part of the EM spectrum.}
\label{fig.1}
\end{center}
\end{figure}
%

The results presented in Fig. 1 are obtained by solving the 
spatial part of the NAE for specified wavelengths of EM waves 
used in the measuring process.  However, the results do not 
take into account the initial state of the electron on its 1s orbital.
The radial probability density for the 1s orbital, ${\cal{P}}_{1s}
(r_a)$, given by the SE is also shown in Fig. 1.  In order to account 
for the electron's initial probablity density on the $1s$ orbital, the 
product ${\mathfrak{P}} (r_a) = {\cal{P}}_{1s} (r_a) \times 
{\cal{P}}_{1} (r_a)$ is calculated and plotted in Fig. 2.  

All the computed probability densities in Fig. 2 are more narrow 
and centered at $r = 2\ a_o$ as compared with those in Fig. 1.  
This is an interesting result as it demonstrates how the original 
electron's probability density on the $1s$ orbital has changed 
because of the measurement process.  Thus, the results of Fig. 
2 are in better agreement with Born's position measurement 
postulates than those given in Fig. 1, which shows that the 
electron's initial probability density calculated by the SE must 
also be accounted for in the procedure that describes the 
measurements.  
 
Since the presented results depend directly on the principal 
quantum number $n$ (see Eqs \ref{eq13} and \ref{eq14}), 
similar calculations can be performed for $n = 2$, $3$, $4$;
however, they will be presented elsewhere.

\section{Experimental verification}

The results presented in Figs 1 and 2 demonstrate that 
the electron originally located on the 1s orbital, with 
its most probable radius at $r = a_o$, is shifted to its 
most probable radius at $r = 2\ a_o$ as a result of the 
measurements.  This means that the electron interacting 
with a measuring apparatus gets confined into an eigenstate 
that corresponds to the measurement.  Comparison of the 
original electron's probability density ${\cal{P}}_{1s} (r_a)$ 
on the orbital $1s$ (see Fig. 1) to ${\mathfrak{P}} (r_a)$ 
plotted in Fig. 2 shows that as a result of the measurement 
the electron is confined to a well-defined position at $r = 2a_o$, 
and that its probability density is very narrow and sharply 
centered at this position, which is consistent with Born's 
position measurement principles (e.g., [41,42]).  The time 
scale for the existence of this eigenstate is given by Eq. 
(\ref{eq7}), which represents its duration. 

The theoretical results obtained in this paper can be 
verified experimentally by using a quantum microscope 
similar to that designed by an international team of 
researchers [54], who used it to measure the orbital 
structure of Stark states in an excited hydrogen atom 
(see their Fig. 3).  They reported a similar trend in 
shifting the maxima and broadening the density 
distributions for shorter wavelengths as those 
shown in Fig. 2.  There are several other measuring 
methods developed to study the quantum-classical 
correspondence [54-59], but some of these methods 
may not be suitable to observe single orbitals in a 
hydrogen atom [58].

After using the NAE to solve the quantum 
measurement problem in this paper, and quantum 
jumps in [47], the NAE has the potential to develop 
new quantum based technologies, which would verify 
its validity and applicability.  As described in a recent
comprehensive review [60], cryptographic systems 
and secure direct communications as well as emerging 
fields like the quantum internet are all based on modern 
quantum technologies. However, to explore potentials of 
developing new quantum technologies based on the NAE 
is out the scope of this paper.

\section{Conclusions}

In this paper, a new asymmetric equation, which 
is complementary to the Schr\"odinger equation,
is used to investigate the measurement problem 
of quantum mechanics.  The obtained results 
demonstrate that while the Schr\"odinger equation 
describes the evolution of the wavefunction prior 
to any measurement, the new asymmetric equation 
represents the behavior of the wavefunction during 
the measurement process.  The temporal solutions 
to the new asymmetric equation give a time-scale
for the duration of the eigenstate that corresponds
to the measurement; probability density at this 
eigenstate is determined from the spatial solutions.  

To describe the transition from unitary (reversible) 
representation of the electron on its $1s$ orbital, 
which is given by the Schr\"odinger equation, to 
its non-unitary (irreversible) evolution resulting 
from the measurement, which is given by the new 
asymmetric equation, the product of both probabilty
densities is calculated.  The main results are that 
the time-scales are or the order of $10^{-17}$ - 
$10^{-15}$ s, depending on the frequency of the 
EM waves used in the measurement, and that the 
computed probability densities are sharply centered 
around $r = 2\ a_o$, which shows that Born's 
position measurement postulates naturally emerge 
from the presented theoretical results.  In other 
words, the new asymmetric equation and first 
principles used to derive it also account for Born's
rules of quantum mechanics. 

The predicted radial probability densities resulting 
from the measurement process can be verified 
experimentally by using a quantum microscope 
(e.g., [54]).  Another way to verify the validity
and applicability of the new asymmetric equation 
would require exploring its potentials in contributing 
to already known quantum based technologies (e.g., 
[60]), or to be used to develop new quantum 
technologies based on this equation; however, 
this is out of the scope of this paper. 

Based on the presented results, it is proposed that 
while the unitary evolution of the wavefunction is 
described by the Schr\"odinger equation, its non-unitary 
behavior caused by measurements is represented 
by the new asymmetric equation; this means that 
both equations are required in nonrelativistic quantum 
mechanics to fully describe the spatial and temporal
evolution of the wavefunction, and the quantum 
measurement problem.\\
%


{\bf Acknowledgments}  I appreciate very much constructive comments 
and insightful suggestions made by three anonymous referees, which 
allowed me to significantly improve this paper.  The author also thanks 
Dora Musielak for reading the earlier version of this manuscript and 
suggesting improvements in its structure and presentation of the results.\\

\end{document}